\documentclass[5p,10pt]{elsarticle}
\usepackage{graphics}
\usepackage{amsmath,amsthm,amsfonts,amssymb}
\usepackage{amssymb}
\usepackage{lineno,hyperref}
\usepackage{multirow}

\journal{Computational Materials Science}
\begin{document}
\begin{frontmatter}

\title{Influence of Lithium Interstitial Doping on the Optoelectronic Properties of NiO and WO$_3$}

\author[label1]{Israel Perez\corref{cor1}}
\cortext[cor1]{Corresponding author}
\ead{israel.perez@uacj.mx}
\author[label2]{Juan Carlos Mart\'inez Faudoa}
\author[label2]{Juan R. Abenuz Acu\~na}
\author[label2]{Jos\'e Trinidad Elizalde Galindo}

\address[label1]{National Council of Science and Technology (CONACYT)-Department of Physics and Mathematics, Institute of Engineering and Technology, Universidad Aut\'onoma de Ciudad Ju\'arez, Av. del Charro 450 Col. Romero Partido, C.P. 32310, Ju\'arez, Chihuahua, M\'exico}
\address[label2]{Department of Physics and Mathematics, Institute of Engineering and Technology, Universidad Aut\'onoma de Ciudad Ju\'arez, Av. del Charro 450 Col. Romero Partido, C.P. 32310, Ju\'arez, Chihuahua, M\'exico}

\begin{abstract}
First principles calculations through density functional theory (DFT)+$U$ method were performed to assess the effect of Li interstitial doping on the optoelectronic properties of NiO and of cubic, hexagonal, and monoclinic WO$_3$. The implications on the structural and electronic properties and their relationship with the optical properties, such as dielectric response and absorption coefficient in the context of electrochromism, are discussed. Li interstitial doping (i.e., electron doping) turns these materials metallic-like even for low Li concentration ($x<0.083$). In metallic-like materials, optical properties are largely influenced by intraband transitions. Compared to NiO, Li-doped NiO shows larger absorption in the visible region while, with respect to WO$_3$, Li-doped WO$_3$ exhibits an increment in absorption in the whole spectrum, particularly, in the infrared region. These results help to understand the changes in optical properties observed in electrochromic devices based on these materials. 
\end{abstract}


\end{frontmatter}

\section{Introduction}
Electrochromic materials change their optical properties by insertion/extraction of charges. These changes can be used to develop electrochromic devices (ECD) with diverse applications such as touch screens, sunglasses, smart windows, among others \cite{wwu18a}. Electrochromic devices are constructed of at least five layers in a pile arrangement supported by a transparent substrate. The lowest and topmost layers are composed of transparent conductors that are used as electrical contacts. The second and fourth layers are made of electrochromic electrodes and in the middle an either solid or liquid/gel electrolyte working as ionic conductor. An applied electric field drives ions into the electrode matrix, resulting in changes in optical properties \cite{pyang16a}.

The electrochromic layers can be composed of viologens, conducting polymers, metal coordination complexes, or transition metal oxides (TMO) \cite{wwu18a,pmsmonk07a}. Among these, TMO have reached a prominent place due to their superior electrochromic properties. Typical TMO used in the fabrication of ECD are WO$_3$, MoO$_3$, TiO$_2$, Nb$_2$O$_5$, NiO, IrO, V$_2$O$_5$, FeO, CoO, MnO, and Ta$_2$O$_5$. Their general structure can be understood in terms of metal-O$_6$ octahedra arranged in corner- and edge-sharing configurations \cite{pmsmonk07a}. Tungsten, molybdenum, nickel, and iridium oxides exhibit the largest coloration efficiencies. From these, WO$_3$ is one of the most studied oxides with a band gap of 2.6 eV and displaying the largest coloration efficiency known so far ($CE$=118 cm$^2$/A$\cdot$s at 633 nm) \cite{pmsmonk07a}. NiO is another material that has attracted the attention of the scientific community due to its low cost, null toxicity, and respectable coloration efficiency ($CE$=42 cm$^2$/A$\cdot$s at 550 nm) \cite{gatak17a,lliu19a}. This compound has a wide band gap around 4.0 eV, is crystalline with cubic structure, and behaves as a p-type semiconductor. Electrochromic layers of these materials can be grown with several deposition techniques such as chemical bath deposition \cite{mahurtado08,xhxia08}, spin coating \cite{ista14,aaghmadi09}, sol-gel method \cite{amsolei13,amsolei12}, pulsed laser deposition \cite{yhe19a}, spray pyrolysis \cite{khabass15,raismail13}, electron beam evaporation \cite{hmoulki14a,spereira14a}, and radio frequency and direct current magnetron sputtering \cite{yabe12,hlchen12,mguziewicz12,amreddy11}. ECD based on WO$_3$ and NiO films combined with other TMO can exhibit higher electrochromic properties than those of the individual oxides, making them of the utmost importance for the fabrication of ECD \cite{ganiklasson07}.

The improvements in the development of ECD have been mainly based on heuristic methodologies and a myriad of clever ideas \cite{wwu18a,granqvist18}. Surprisingly, the changes in the optical properties are still not well understood. The understanding of electrochromism in WO$_3$ has been developed along two lines of thought. The most popular approach is based on polaron theory. In this case, the change in the optical properties is caused by photon-induced polaron hops between tungsten cations in various oxidation states. The mobility and formation of polarons in the presence of both oxygen vacancies and lithium intercalation for the cubic and monoclinic structures of WO$_3$ have been investigated \cite{ebroclwik06a,nbondarenko15}. This mechanism explains well the infrared absorption but fails in the visible and ultraviolet regions \cite{granqvist93}. Moreover, it has been shown that small polarons in these crystalline phases are unstable \cite{wwang16}. The other approach addresses the phenomenon in terms of band theory \cite{cggranqvist95}. It is assumed that electrons injected through a cathodic electrode occupy new levels at the bottom of the conduction band. These electrons fill the empty states in this band and optical absorption can be seen as a transition between unoccupied and occupied states in the conduction band. Most work following this path has been focused on studying the structural, electronic, and optical properties for the cubic and hexagonal phases using density functional theory (DFT) \cite{ahjelm96a,bingham05,amahmoudi16a,cyang14,cyang16}. The results shed some light on the effect of Li intercalation on the optical properties of WO$_3$, however these crystalline structures are not the most common structures found in ECD. During the construction of ECD, WO$_3$ is usually deposited at room temperature to avoid diffusion effects among layers and therefore the most stable phases are amorphous and monoclinic \cite{gatak17a,xsong15,atak18,qliu18}. The cubic is unstable at normal conditions \cite{crichton03} while the hexagonal phase may show up as a metastable phase or as an admixture of monoclinic and hexagonal phases in thin films \cite{hawriedt89a,tnanba89a,tvogt99a,krlocherer99a}. The effect of alkali-metal insertion on the electronic band gap of monoclinic WO$_3$ was investigated in the context of photocatalysis using DFT with a hybrid functional B3LYP \cite{stosoni14}, unfortunately, the optical properties were not addressed here. Hence, so far, ab-initio investigations on the optical properties have only focused on both the undoped phases \cite{yping13} of monoclinic WO$_3$ and doped phases within the context of polaron theory. In this work we delve into the optical properties of monoclinic WO$_3$ with Li insertion within the context of band theory. 

On the other hand, NiO is a typical strongly correlated material that has been extensively studied in the literature both theoretically and experimentally \cite{vianisimov91a,ftran06a,gasawaztky84,cykuo17a}. The effect of lithium doping on the electronic structure of NiO was investigated with ab-initio methods using several levels of theory such as hybrid functionals, DFT+$U$, and DFT plus dynamical mean field theory \cite{yhe19a,hchen12,wrobel20}. It was found that Li substitutional doping creates new states in the conduction band which in turn impacts the band gap size. Yet, within the context of electrochromism, it is believed that Li is incorporated into the interstices of the matrix \cite{cggranqvist95}, a matter that, to our knowledge, has not been addressed so far in the literature of this compound. Therefore a research tackling this aspect is of great relevance for the development of NiO-based ECD.

To fill these theoretical gaps we perform ab-initio calculations based on DFT for monoclinic WO$_3$ and NiO with and without Li insertion and compare our results with the cubic and hexagonal phases of WO$_3$. Both the density of states and band structure are computed in the context of the DFT+$U$ approximation. To assess the influence of Li insertion on the optical properties of these systems, the dielectric function and absorption coefficient are reckoned. We found that Li insertion not only expands the unit cells but also transforms the materials from a semiconductor to a metallic-like material. Absorption of the undoped systems is enhanced in the infrared and visible regions by Li insertion due to intraband transitions. We think that the findings of this research are valuable for a better understanding of electrochromism that can be helpful for the design of ECD. 

For the purpose of this research, the article is organized as follows. In section \ref{sec2}, the outline of the theoretical methodology is given. Section \ref{sec3} describes the results of the electronic structure through the spin-polarized density of states and band structure, and discusses the optical properties; paying special attention to the dielectric tensor and the absorption coefficient. In section \ref{sec4} we give our conclusions.

\section{Theoretical methods}
\label{sec2}
\subsection{Computational Methodology}
\begin{figure}[t!]
\begin{center}
\includegraphics[width=8.7cm]{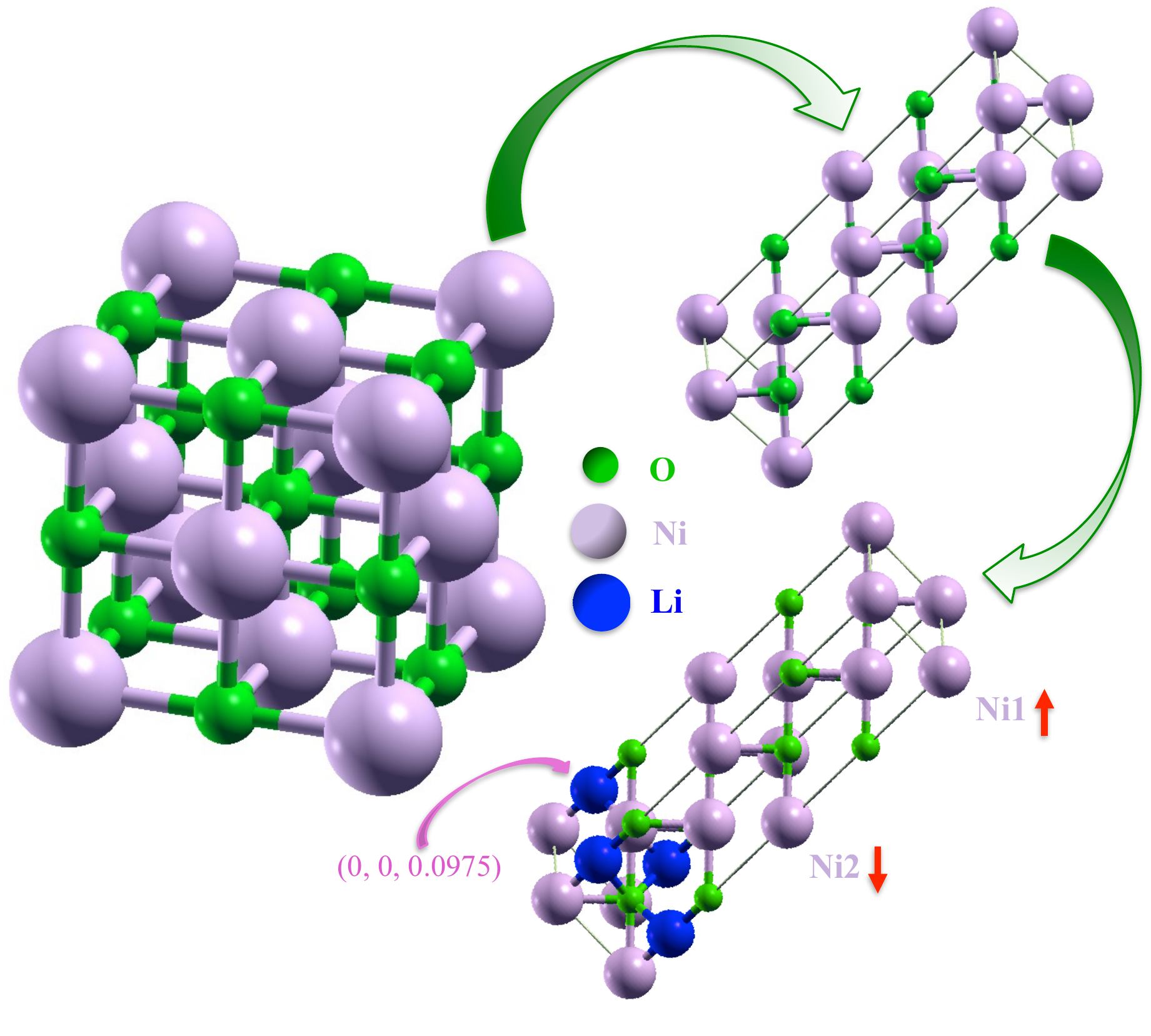}
\caption{Crystal structures of NiO. Cubic (left), pristine antiferromagnetic (right-upper) and Li-doped antiferromagnetic (right-lower). Red arrows indicate spin-polarization orientation for each nickel atom. The position of the dopant is also given. Atomic structure visualization provided by
the Xcrysden software package \cite{kokalj99}}
\label{NiO}
\end{center}
\end{figure}

NiO can crystallize in rhombohedral, monoclinic, and cubic structures \cite{neumann84,gaslack60,yshimomura56}; however, in the fabrication of ECD only the cubic phase has been reported in the literature \cite{pmsmonk07a,gatak17a,hmoulki14a,spereira14a} and for this reason we constraint ourselves to discuss this phase. The cubic phase of NiO is antiferromagnetic (AFM) with a rock-salt structure and space group $Fm\bar{3}m$ but taking into account the antiferromagnetic order along the [111] direction, the symmetry is lowered to rhombohedral one with space group $R\bar{3}m$. This leaves two inequivalent atoms, Ni1 and Ni2, for spin-polarized calculations. The spin orientation for these atoms is set to 'up' and 'down', respectively (see Figure \ref{NiO}). 

\begin{figure}[t!]
\begin{center}
 \includegraphics[width=9cm]{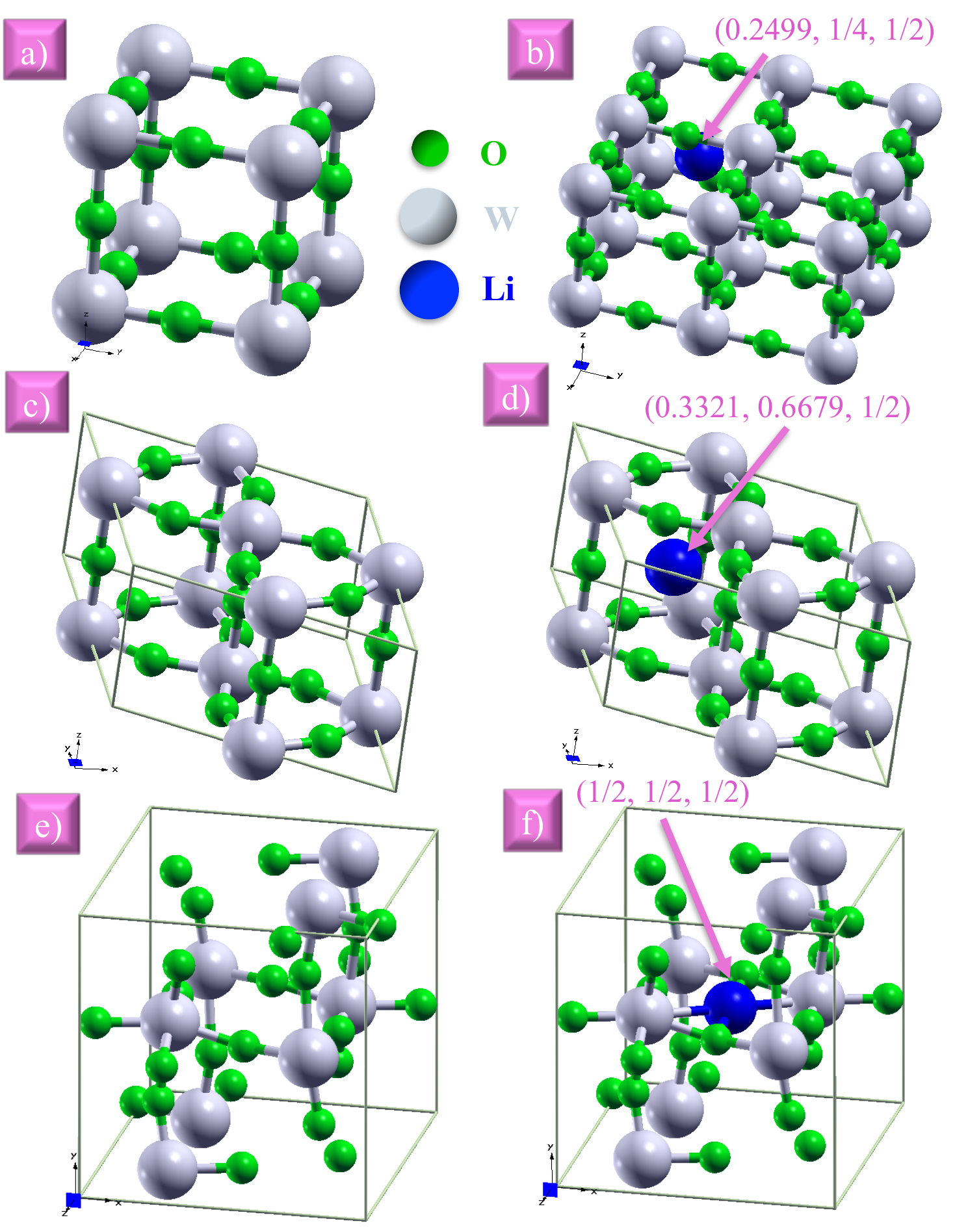} 
\caption{Crystal structures of WO$_3$: cubic (a,b); hexagonal (c,d), and monoclinic (e,f). Left column: pristine structures. Right column: Li-doped structures. The position of the dopant within the supercell is indicated in parenthesis. Atomic structure visualization provided by
the Xcrysden software package \cite{kokalj99}}
\label{WO3struct}
\end{center}
\end{figure}

WO$_3$ exhibits polymorphism as a function of temperature, however, to avoid diffusion effects, WO$_3$ in ECD is deposited at room temperature. At this temperature the most stable crystalline phase is $\gamma-$WO$_3$ whose symmetry is $P2_1/n$. Another room temperature phase of WO$_3$ is a metastable phase that crystallizes in hexagonal structure with symmetry group $P6/mmm$. This structure commonly appears in thin films and therefore is of appreciable interest for this research. As a reference we also consider the cubic phase (see Figure \ref{WO3struct}). 

Intercalation of alkali metals in TMO leads to complicated structural transitions that largely depend on the dopant concentration $x$. For instance, during Li insertion in WO$_3$, the cell expands, while, during Li extraction, the cell contracts. It has been reported that for $x> 0.1$, Li$_x$WO$_3$ undergoes a sequence of structural transitions, namely: monoclinic$\rightarrow$tetragonal$\rightarrow$cubic \cite{pmsmonk07a,qzhong92,bingham05}. Thus, to avoid structural transitions and study the effect of Li insertion on the optical properties of the four systems, we keep $x<0.083$ by adding one single lithium atom to the parent lattice. For the simulation of Li in the interstices of NiO and the hexagonal structure of WO$_3$, we use a $2\times1\times1$ supercell with 12 ($x=0.083$) and 16 atoms ($x=0.062$), respectively. For the cubic phase of WO$_3$ we use a $2\times2\times1$ supercell with 16 atoms ($x=0.062$), and for the monoclinic structure a 32-atom cell ($x=0.031$). For the sake of simplicity we shall call the doped systems as: LiNiO and cubic, hexagonal, and monoclinic LiWO$_3$. 

Calculations are carried out within the context of DFT using the full-potential linearized augmented plane wave plus localized orbitals (FP-LAPW+lo) method as implemented in \textsc{wien2k} code \cite{pblaha20a}. For the exchange correlation functional, we employ the Perdew-Burke-Ernzerhof approximation \cite{jpperdew96a} alongside a self-interaction correction variant in the generalized gradient approximation (GGA)+$U$ for both cases: undoped and doped systems. The effective potential $U_{\textrm{eff}}=U-J$, where $U$ is the on-site Hubbard parameter and $J$ the Hund coupling, is applied to the Ni $3d$ and W $5d$ states. The values of $U_{\textrm{eff}}$ for the systems are assigned as follows. NiO is a well-known system whose typical value is 7 eV and this same value was used for LiNiO \cite{vianisimov91a,ftran06a}. For the pure systems of WO$_3$ we employ the method adopted by Bondarenko et al. \cite{bondarenko14}. We vary the value of $U_{\textrm{eff}}$ from 0 eV to 14 eV and seek for the values that better account for the equilibrium structural parameters and the band gap. These same values are used for the corresponding doped systems (see Table \ref{rmtu}). The reciprocal space integrations for the pure systems are realized with 1000 k-points in the full Brillouin zone (FBZ) using the tetrahedron method. For the doped systems we used 100, 120, 250, and 100 k-points in the FBZ for LiNiO, cubic, hexagonal, and monoclinic LiWO$_3$, respectively. The cut-off energy between core and valence states is set to $-6$ Ry for all systems. The radii of the muffin tin (R$_{MT}$) spheres for the atoms are chosen so that the neighbouring spheres are nearly touching. The selection of these radii allows us to avoid any charge leakage (see Table \ref{rmtu}). 
 \begin{table}[b!]
  \centering 
  \caption{Parameters for the systems. Crystal system, space group, radius of the muffin tin sphere ($R_{MT}$ in atomic units) and effective potential ($U_{\text{eff}}$ in eV).}
  \label{rmtu}
  \begin{tabular}{cc|cccc|c}
\hline \hline
System & Space &   \multicolumn{4}{c|}{$R_{MT}$} & $U_{\textrm{eff}}$ \\
  & Group  & O & Ni & W & Li  & \\ \hline 
 NiO& $R\bar{3}m$   &  1.7 & 1.9 & ---& ---  & 7.0  \\ 
 LiNiO  & $R\bar{3}m$ & 1.7 & 1.9 & --- & 1.35   & 7.0 \\ \hline 
 WO$_3$& $Pm\bar{3}m$   & 1.5 & ---  & 1.7 & ---  & 12.0   \\
  LiWO$_3$  & $Pm\bar{3}m$  & 1.5 &---  & 1.7  & 1.35 & 12.0\\
 \hline 
  WO$_3$  & $P6/mmm$  & 1.5 &--- & 1.7 & --- & 7.0 \\
  LiWO$_3$  & $P6/mmm$  & 1.5 & --- & 1.7 & 1.35 & 7.0 \\ \hline
 WO$_3$  & $P2_1/n$  & 1.5 &--- & 1.7  & ---  & 7.0\\ 
  LiWO$_3$  &$P2_1/n$  & 1.5 &--- & 1.7 & 1.35 & 7.0 \\ \hline 
\end{tabular}
\end{table}

For the expansion of the basis set, we set the product of the smallest of the atomic sphere radii $R_{MT}$ and the plane wave cutoff parameter $K_{max}$ to $RK_{max}$= 7. During the self-consistent field cycle the energy convergence is less than 100 $\mu$Ry/unit cell with a force convergence of 1 mRy/Bohr. For higher resolution in the computation of both electronic and optical properties, the size of the k-point grid is increased. For pure monoclinic WO$_3$ the grid is set to 4000 k-points and to 8000 k-points for the rest of the pristine systems. For the Li-doped systems we use grids of 500 k-points, for all systems.

Lithium positions are chosen based on several considerations. Firstly, large enough cavities to allocate the Li atom must be available in the unit cell. The cavity size should guarantee no overlapping of atomic spheres (and thus charge leaking) since the R$_{MT}$ values are critical, particularly during the optimization process. Secondly, the incorporation of the atom should respect the lattice symmetry and structural stability. This is achieved by keeping the doping concentration relatively low \cite{qzhong92,bingham05}. The structural stability is assessed by computing the enthalpy of formation which, for the ground state at zero pressure ignoring zero-point contributions, is equal to the energy of formation \cite{amahmoudi16a}. These conditions greatly reduce the number of available sites for insertion. 

The larger cavities for the cubic and monoclinic phases of WO$_3$ are found at the center of the cell and at the center of the six faces. The hexagonal structure can host large metal atoms in hexagonal and trigonal cavities. For the sake of analysis, we carry out the calculations placing the Li atom in only one of the several large cavities for each system (note that the values reported are those after optimization). In the case of LiNiO, the Li atom is placed at (0, 0, 0.0975) in the rhombohedral cell. For the cubic phase of LiWO$_3$ the atom is incorporated at the center of one of the supercell units at (0.2499, 1/4, 1/2). For the hexagonal phase of LiWO$_3$ we used a trigonal cavity in the plane of the apical oxygen atoms at (0.3321, 0.6679, 1/2). In the case of the monoclinic phase of LiWO$_3$, the dopant is located at the center of the cell (see Figure \ref{NiO} and \ref{WO3struct} for illustration). The initial lattice parameters for the pure structures are taken from the literature \cite{hawriedt89a,tnanba89a,tvogt99a,krlocherer99a,wyckoff63}. For structure relaxation, we first carry out lattice optimization and then the internal coordinates are relaxed. We repeat this procedure twice and check convergence. All atoms are fully relaxed up to 1 mRy/Bohr Hellman-Feynman forces. The optimized lattice parameters are given in Table \ref{tablestruct}.

\subsection{Optical Theory}
\label{optteo}
The understanding of the optical properties is vital for the development of absorbing and reflective electrochromic devices. We have used the electronic structure calculations to compute the complex dielectric tensor $\varepsilon_{ij}$ and the absorption coefficient $\alpha$, using the program \textsc{optic} implemented in the \textsc{wien2k} code. The theory departs from the Linhard dielectric function in the random phase approximation \cite{cambrosch06a}:
\begin{equation}
\varepsilon(q,\omega)=1+\frac{2\Phi(q)}{\Omega_c}\sum_k\frac{f_l(\epsilon_{k+q})-f_l(\epsilon_{k})}{\epsilon_{k+q}-\epsilon_{k}-\hbar \omega};
\label{dieconst1}
\end{equation}
where $\Phi(q)=\frac{4\pi e^2}{q^2}$ is the Coulomb interaction, $q$ is the wavenumber vector, $\omega$ is the angular frequency, $\Omega_c$ the unit cell volume, $f_l$ the Fermi distribution function, $\epsilon_k$ the single particle energy and $\hbar$ is the reduced Planck constant. The 2 comes from the summation over spins. The dielectric function in the lattice space becomes a complex symmetric tensor $\varepsilon_{ij}(\omega)$ with contributions from both intraband and interband transitions (It is stressed that in this theory only direct transitions are taken into account). From the imaginary part [$\varepsilon_2(\omega)$] the corresponding real part [$\varepsilon_1(\omega)$] can be obtained by the Kramers-Kronig relations
\begin{equation}
 \varepsilon_1(\omega)=1+\frac{2}{\pi}\mathcal{P}\int_0^\infty\frac{\omega'\varepsilon_2(\omega')}{\omega'^2-\omega^2}d\omega',
\label{kkr}
\end{equation}
where $\mathcal{P}$ indicates the principal part of the integral. With this information one can compute all other optical properties. The reflectivity at normal incidence is given by
\begin{equation}
R(\omega)=\frac{(\textrm{n}-1)^2+\textrm{k}^2}{(\textrm{n}+1)^2+\textrm{k}^2};
\label{rii}
\end{equation}
with n and k being the real and imaginary part of the complex refractive index ($\tilde{\textrm{n}}$=n$+i$k) whose components are the refractive index and extinction coefficient given by:
\begin{equation}
\textrm{n}(\omega)=\sqrt{\frac{|\varepsilon(\omega)|+\varepsilon_1(\omega)}{2}},
\label{nii}
\end{equation}
\begin{equation}
\textrm{k}(\omega)=\sqrt{\frac{|\varepsilon(\omega)|-\varepsilon_1(\omega)}{2}}.
\label{kii}
\end{equation}
The absorption coefficient is related to the extinction coefficient by the following relation:
\begin{equation}
\alpha(\omega)=\frac{2\omega \textrm{k}(\omega)}{c}.
\label{alphaii}
\end{equation}
This parameter plays a key role in understanding the relationship between the electronic structure and the optical properties as we discuss below. Lastly, the effective number of electrons per atom $N_{eff}$ partaking in the absorption process as a function of energy can be determined from the following sum rule:
\begin{equation}
\int_0^{\omega'}\omega \varepsilon_2(\omega)  d\omega =N_{eff}(\omega').
\label{sumr}
\end{equation}

\section{Results and Discussion}
\label{sec3}

\subsection{Atomic Structure}
To check structure stability, the formation energies per atom for the doped systems are reckoned. The results for LiNiO, cubic, hexagonal, and monoclinic LiWO$_3$ are: $-13.255$ eV, $-9.375$ eV, $-3.733$ eV, and $-5.398$ eV, respectively. These findings indicate that all structures are stable. We stress that small variations in position of the dopant atom do not affect significantly the energy of formation.
\begin{table*}[t!]
  \begin{center}
    \caption{Cell parameters (in \AA$\;$ and $^{\circ}$) compared to other experimental and calculated results. Values reported from references \cite{qzhong92,nbondarenko15,ahjelm96a,bingham05,cyang14,stosoni14,hchen12,ddong18}. Parameters are given after structure relaxation.}
    \label{tablestruct}
    \begin{tabular}{ccccc}
      \hline \hline
      System & Lattice &This & Experimental [Ref.]& Calculated [Ref.] \\
        & Parameter &Work & &\\ \hline \hline
      NiO & $a$ & 4.1933 & 4.18 \cite{hchen12}&4.20 \cite{hchen12}\\  \hline
       LiNiO & $a$ & 4.3935 & 4.236 (4\% Li) \cite{ddong18} &---\\ \hline
      Cubic WO$_3$ & $a$ & 3.7327 &3.78 \cite{ahjelm96a}, 3.8144  \cite{bingham05} &3.84 \cite{ahjelm96a}, 3.74   \cite{bingham05}\\   \hline
      Cubic LiWO$_3$  & $a$ & 3.749& 3.729 \cite{qzhong92}, 3.71 \cite{ahjelm96a} &3.88 \cite{ahjelm96a} \\  \hline
  Hexagonal        & $a$ & 7.2955 & 7.298 \cite{bingham05} & 7.4103 \cite{bingham05}, 7.453 \cite{cyang14} \\ 
       WO$_3$  & $c$ & 3.8976 & 3.899 &3.8144 \cite{bingham05}, 3.833  \cite{cyang14}  \\ \hline
  Hexagonal     & $a$ & 7.4619 &7.405 \cite{bingham05} & 7.4007 \cite{bingham05}, 7.471 \cite{cyang14}\\ 
    LiWO$_3$    & $c$ & 3.9866  & 3.777 & 3.8219 \cite{bingham05}, 3.841 \cite{cyang14} \\ \hline  
             & $a$ & 7.3035 &7.306 \cite{qzhong92} & 7.34 \cite{nbondarenko15}, 7.44 \cite{stosoni14}\\ 
   Monoclinic   & $b$ & 7.5374 & 7.540 & 7.58 \cite{nbondarenko15}, 7.73 \cite{stosoni14}\\ 
     WO$_3$    & $c$ & 7.6893 &7.692 & 7.73 \cite{nbondarenko15}, 7.91 \cite{stosoni14} \\  
         & $\beta$ & 90.88 & 90.881 & 90.89 \cite{nbondarenko15}, 90.2 \cite{stosoni14}\\  \hline
     & $a$ & 7.4701 & 7.310 \cite{qzhong92} &  7.46 \cite{stosoni14} \\ 
      Monoclinic & $b$ & 7.7093 & 7.540 &7.62 \\ 
     LiWO$_3$  & $c$ & 7.8648 & 7.695 & 7.96 \\ 
         & $\beta$ & 90.881 & 90.881& 90.3\\  \hline  \hline
    \end{tabular}
  \end{center}
\end{table*}

Table \ref{tablestruct} shows the results for the relaxed parameters of our systems in comparison to previous experimental and theoretical work. For the case of LiNiO, the cell size increased about 5\% with respect to the undoped oxide. It is worth noting that the value of our lattice parameter is larger than the experimental one \cite{ddong18}. This can be attributed to the difference in doping levels. In the experimental report, the lithium concentration is 4\% while in this work a concentration of 8.3\% is used, which is more than double. Since, the concentration is larger one would expect a larger cell. For WO$_3$, one observes, in all cases, as expected, an increase in cell size of 4\%, 2.3 \%, and 2.3\% for cubic, hexagonal, and monoclinic, respectively. If we compare our results with those in the literature, we note that for the cubic phase there are discrepancies in both the experimental and theoretical values, although they are comparable \cite{qzhong92,ahjelm96a,bingham05}. The agreement of our results with experimental values is excellent for the pure hexagonal and monoclinic phases \cite{qzhong92,bingham05}; while other theoretical works found values slightly higher than ours \cite{ahjelm96a,bingham05,cyang14,stosoni14}. For the doped hexagonal and monoclinic phases, the agreement is fair with one theoretical report \cite{cyang14,stosoni14}, although slightly larger than the experimental value \cite{qzhong92}.

\subsection{Electronic properties of NiO and LiNiO}
The electrochromic properties in ECD are determined by the transport and optical properties of the electrochromic material \cite{gatak17a,cggranqvist95}. In particular, the colouring and bleaching processes are linked to the absorption of the TMO which strongly depends on the magnitude of the energy gap. From Table \ref{tablestruct} we just have concluded that the lattice parameters for the doped structures are larger than those for the pristine structures. This indicates that Li insertion exerts pressure on the initial atomic positions, deforming and expanding the lattice; such structural changes then affect the electronic structure and consequently the optical response of the systems \cite{bingham05,stosoni14}. It is therefore important to delve into the electronic properties and analyze how the optical properties are influenced by Li intercalation.

\begin{figure}[t!]
\begin{center}
\includegraphics[width=9cm]{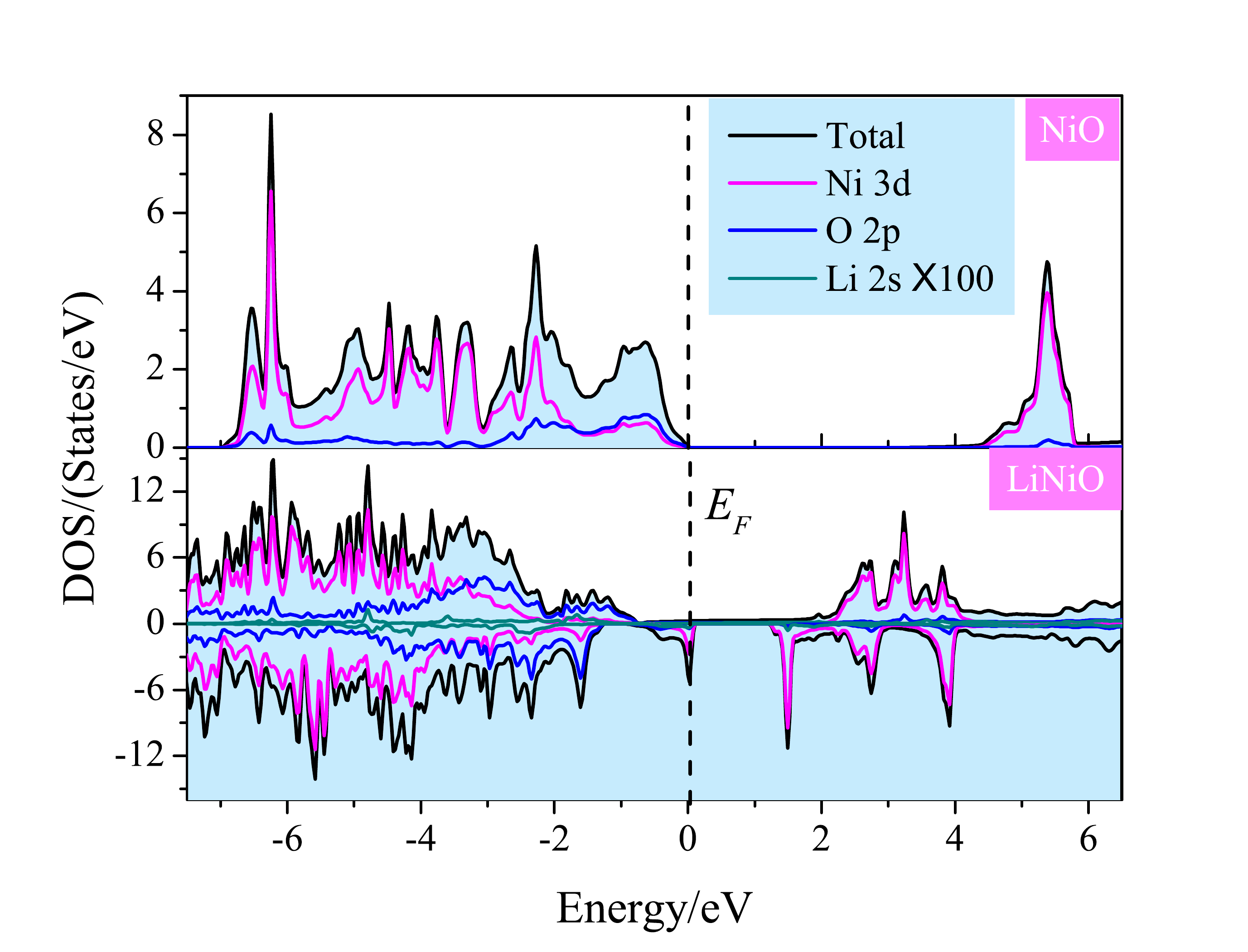}
\caption{Total and projected density of states of NiO and LiNiO. Vertical dashed line represents the Fermi level}
\label{LINIONIO}
\end{center}
\end{figure}

NiO is considered a classical Mott insulator with localized $3d$ electrons causing strong correlation effects. It can be hole doped for instance through substitutional doping or Ni deficiency and it can be electron doped by interstitial doping usually with an alkali metal. Figure \ref{LINIONIO} shows the total and partial density of states (DOS) of NiO and LiNiO. The Fermi level ($E_F$) is denoted by the vertical dashed line that separates the valence band (VB) from the conduction band (CB). The DOS of NiO for both spin channels is symmetric and for this reason the DOS for minority spin is not shown. Accordingly, the band gap ($E_g$) for NiO is about 3.4 eV slightly lower than the experimental value (3.8-4.2 eV) \cite{gasawaztky84}. The whole spectrum is mainly dominated by Ni $3d$ states, except for the interval from -2 eV to $E_F$ where the O $2p$ states are more intense. Ni $4s$ states have a negligible contribution and are not shown. The prominent feature observed in the CB is located around 5 eV above the Fermi level and is dominated by Ni $3d$ states. 

When Li is present, the symmetry for both spin channels is broken, turning the material a half-metallic semimetal with literally no band gap at the Fermi level \cite{coey04,katsnelson08}. There is an increment in the number of states beyond the Fermi level for both spin channels and 1.1 eV gap is seen only for the spin-down channel, 0.1 eV above the Fermi level. The contribution to the total DOS comes mainly from Ni $3d$ and O $2p$ states (4$s$ states are not shown due to imperceptible contribution). The latter  dominate the regions from 4.3 eV to higher energies for both spin channels, from $-3.5$ eV to $2$ eV for the spin-up channel, and from $-3.5$ to the Fermi level for the spin-down channel. For the rest of the spectrum both spin channels exhibit Ni $3d$ character. Note that, in comparison to the DOS of NiO, the prominence of the O $2p$ states extends roughly 1.5 eV to lower energies. The contribution of Li $2s$ states is negligible and was augmented 100 times for visualization. In spite of their low occupation, it is clear that the incorporation of Li in the lattice has a significant impact on the electronic structure.

\begin{figure}[t!]
\begin{center}
\includegraphics[width=7cm]{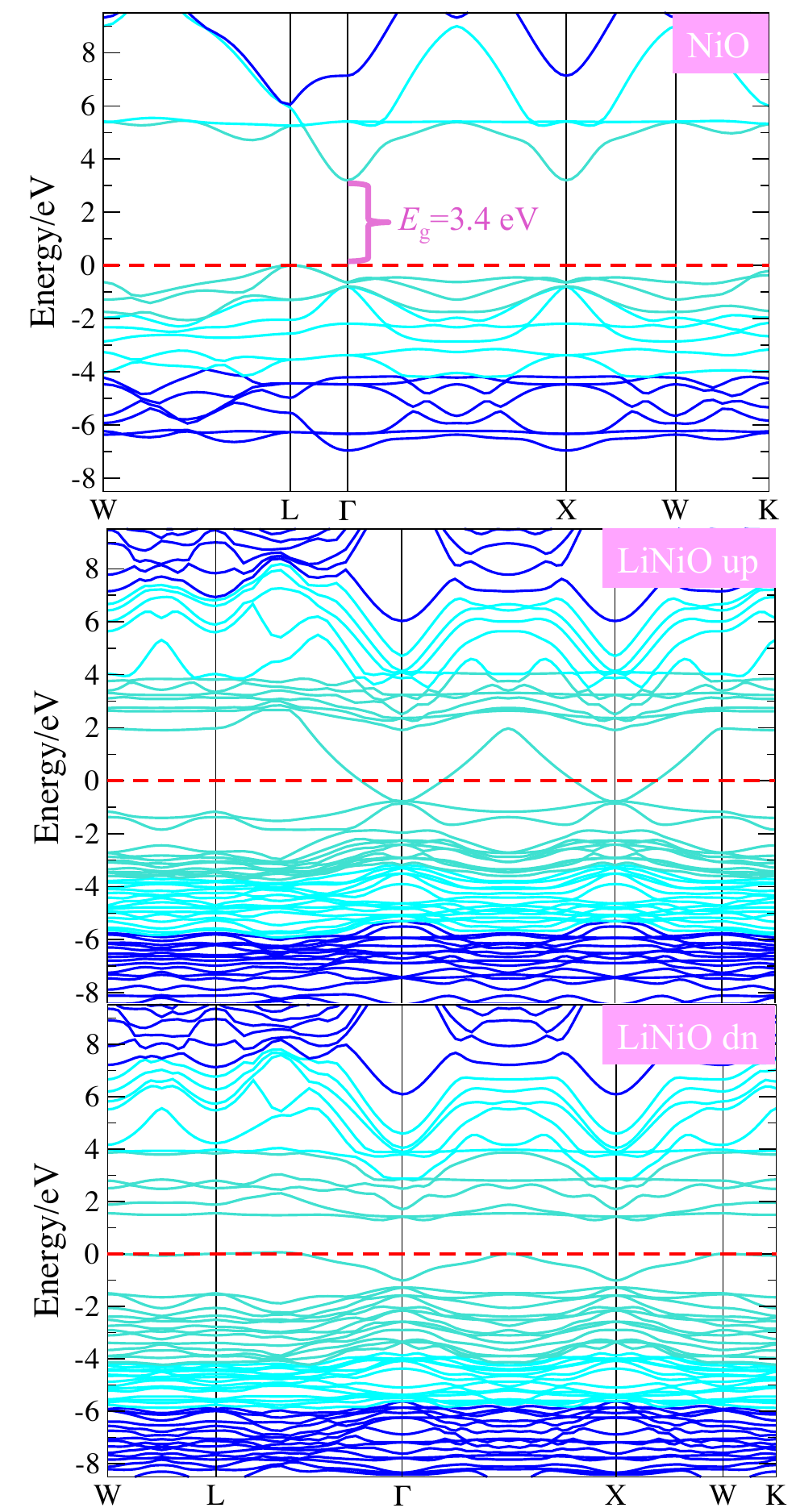}
\caption{Band structure of NiO and of LiNiO for majority (up) and minority (dn) spin channels. Red dashed lines represent the Fermi level.}
\label{bsnio}
\end{center}
\end{figure}

The band structure of both NiO and LiNiO for majority (up) and minority (dn) spin channels is shown in Figure \ref{bsnio}. The band structure of NiO compares well with previous studies \cite{vianisimov91a,ftran06a}. The VB maximum (VBM) is seen at the $L$ point while the CB minimum (CBM) is at the $\Gamma$ point, showing that this material is an indirect band gap semiconductor. When Li is inserted, the symmetry between the spin channels is broken revealing a band structure typical of a half-metallic semimetal. In an earlier work, ab-initio calculations of  Li substitutional doping in NiO for several concentrations ($0.0125<x<0.25$) were carried out \cite{yhe19a,hchen12} and an asymmetry in both spin channels was also reported. As the Li (or hole) concentration was increased, the number of states above the Fermi level also increased, narrowing the energy gap. 
For $x=0.25$, $E_g$ was heavily reduced to 0.3 eV. These results showed that by varying Li concentration, the size of the band gap can be modulated. In the present work, however, Li interstitial doping turns the material metallic-like for $x=0.083$. This finding marks a significant difference between the effects produced by substitutional and interstitial doping. In the former, a Ni$^{2+}$ cation is replaced by a Li$^{+1}$ cation which creates a hole (i.e. hole doping) and unoccupied levels of Ni are created above the Fermi level; this then reduces the size of the gap. In interstitial doping, electrons are added to the system (i.e. electron doping), this rises the Fermi level to the CB and creates occupied levels below the Fermi level, collapsing the gap.  

\subsection{Electronic properties of WO$_3$ and LiWO$_3$}
\begin{figure*}[t!]
\begin{center}
\includegraphics[width=15cm]{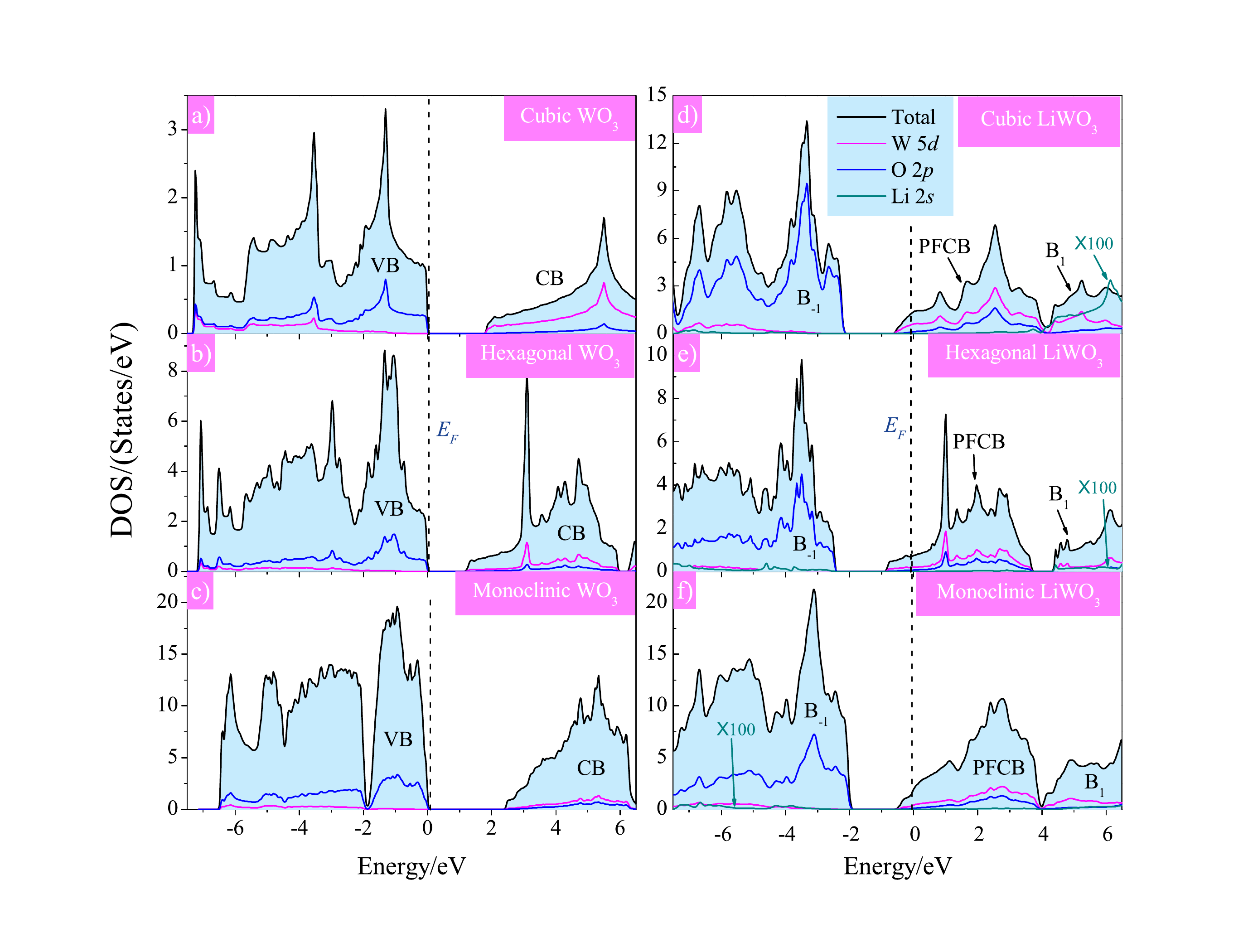}
\caption{Total and projected density of states of WO$_3$ (a-c) and LiWO$_3$ (d-f). Vertical dashed line represents the Fermi level. For band nomenclature see text.}
\label{dosWO3}
\end{center}
\end{figure*}
In Figure \ref{dosWO3} the total and projected DOS of WO$_3$ and LiWO$_3$ are plotted for the three crystalline structures. The dashed line also represents the Fermi level. Since all structures show a symmetric DOS for both spin channels, only one spin channel is presented. The pristine structures exhibit an energy gap that depends on the crystal symmetry. The calculated values of $E_g$ are (1.8, 1.2, 2.5) eV for cubic, hexagonal, and monoclinic, respectively. The latter is close to experimental data (2.6 eV-3.2) eV \cite{cggranqvist95,zheng11,kleperis97,koffyberg79,fzhu19}. The band gap for cubic and hexagonal structures is underestimated although it is larger than other theoretical reports that did not incorporate the self-interaction correction \cite{ahjelm96a, bingham05,amahmoudi16a}. In all cases the CB and the VB are formed by hybridization of W $5d$ and O $2p$ states, but the CB is mainly dominated by W $d$ states while the VB has O $p$ character.

Upon Li insertion all systems display a metallic character, the spectrum changes drastically shifting the Fermi level to the CB and slightly reducing the size of the band gap: an effect known as Burstein-Moss shift. This effect is typical of degenerate n-type semiconductors when the doping level is high such that the semiconductor turns into a metal. Incidentally, there is no symmetry breaking between spin up and spin down channels. At this point it is worth making a crucial clarification that will be useful for our future discussion. In semiconductors the denotations VB and CB are well understood but in metals this terminology is not appropriate since the Fermi level is in the CB which has become a partially filled band. 
 \begin{figure*}[t!]
\begin{center}
\includegraphics[width=17.5cm]{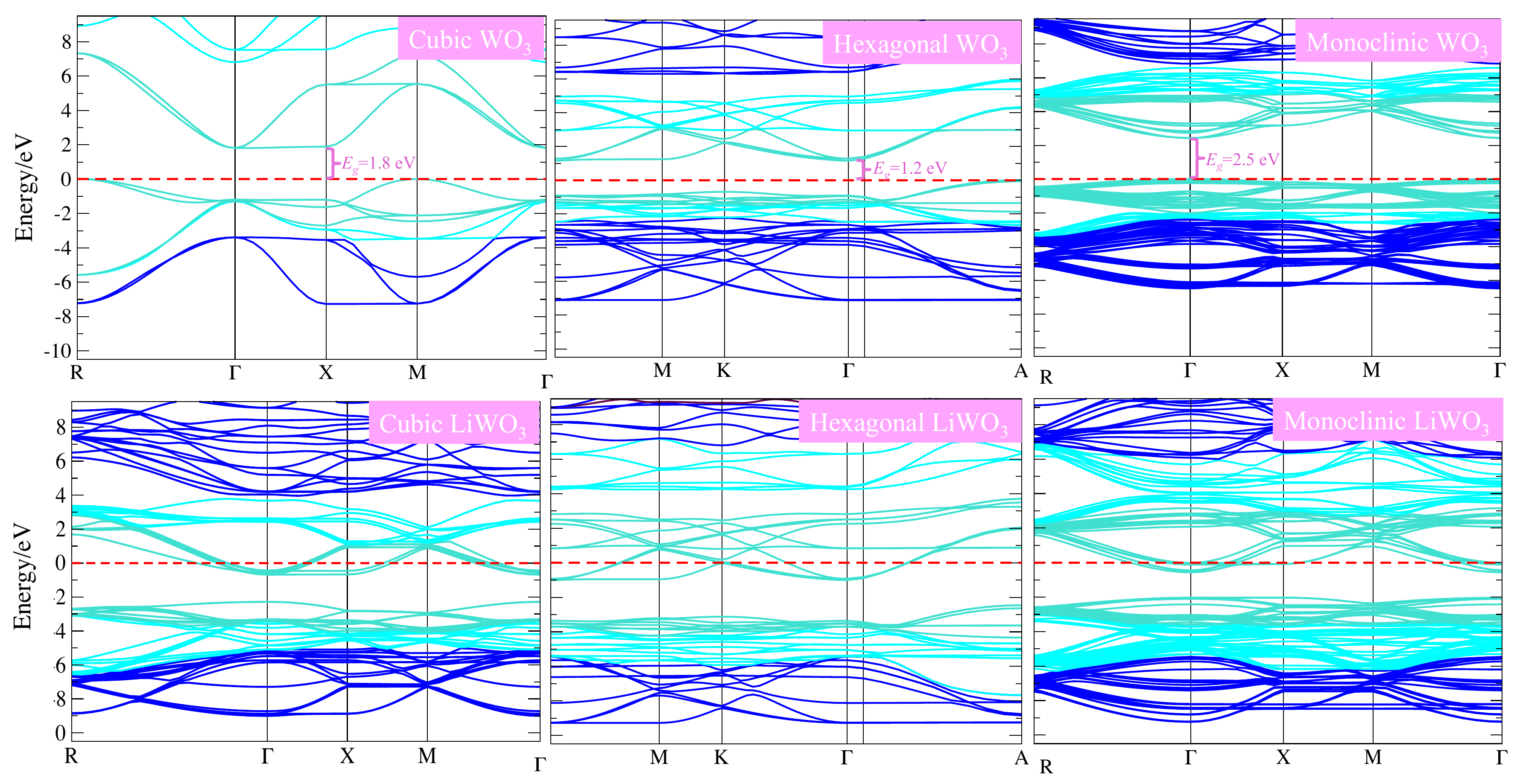}
\caption{Band structure of WO$_3$ and LiWO$_3$ for the different structures. Red dashed lines represent the Fermi level.}
\label{bswo3}
\end{center}
\end{figure*}
Following the work of Xu et al. \cite{xu12a}, it is therefore more convenient to rename the bands for the doped systems: the CB is now termed the partially filled CB (PFCB), the band below this one is the highest fully occupied band (B$_{-1}$), and the band above is the lowest unoccupied band ($B_1$). Under this nomenclature the W $5d$ states dominate both the PFCB and the $B_1$, whereas the B$_{-1}$ presents O $2p$ character and a remarkable increment, compared to the pure phases, in the number of  states. For the sake of visualization, the Li $2s$ states have been augmented 100 times. These states lay beyond the Fermi level higher than the W $5d$ states. This can be explained if we keep in mind that the $2s$ level of atomic Li is higher in energy than the $5d$ level of atomic W and thus one would expect that the W $5d$ states lay closer to the Fermi level. The position of the Li $2s$ states also helps us elucidate the role played by the Li atom. Since such states lay beyond the Fermi level the corresponding valence electrons enter the PFCB (more precisely the $t_{2g}$ conduction band) thus leaving the Li atom in an ionized state \cite{bullett83}. This scenario reveals that Li insertion provides itinerant electrons to the system, in agreement with the metallic character reported in experiments \cite{green96,kyoshimatsu17}. Therefore, the transition to metallic is anew due to the injection of electrons to the system (electron doping) \cite{fzhu19,wwang17,walkingshaw04}.

The band structure for WO$_3$ and LiWO$_3$ is presented in Figure \ref{bswo3}. Starting with the pure phases we see that the monoclinic phase shows more flat bands in comparison to the cubic and hexagonal phases in agreement with the shapes of the corresponding DOS. One of the most conspicuous features spotted in these graphs is the different nature of the band gap. The CBM for both the cubic and hexagonal structures is at the $\Gamma$-point and the VBM is at the M-point and A-point, respectively; making these materials indirect band gap semiconductors. On the other hand, the VBM and the CBM for the monoclinic phase are at the $\Gamma$-point, revealing an direct band gap. Since the same exchange-correlation functional was used for the three structures, such differences can only be attributed to the crystal symmetry. Bear in mind that the cubic symmetry is not found at normal conditions, however, the hexagonal and monoclinic are routinely obtained in the fabrication of ECD. From this standpoint, the nature of the fundamental gap has not been resolved \cite{zheng11,kleperis97,koffyberg79,green93}. As we discuss below, the monoclinic phase is highly anisotropic and the optical properties greatly vary in different directions which makes this material difficult to characterize. On the theoretical side, the use of different functionals in the monoclinic phase has led to dissimilar predictions although the energy difference between direct and indirect gaps is only 0.05 eV \cite{yping13,fwang11,johansson13a}.  

Insertion of Li modifies the band distribution and a considerable amount of new states emerge around the Fermi level. The $B_{-1}$ is flat and the band gaps slightly narrow. Similar to the case of LiNiO, this is due to the fact that the $2s$ level of atomic lithium is higher in energy than the $5d$ level of atomic tungsten.

\subsection{Optical Properties}
\begin{figure*}[t!]
\begin{center}
\includegraphics[width=16cm]{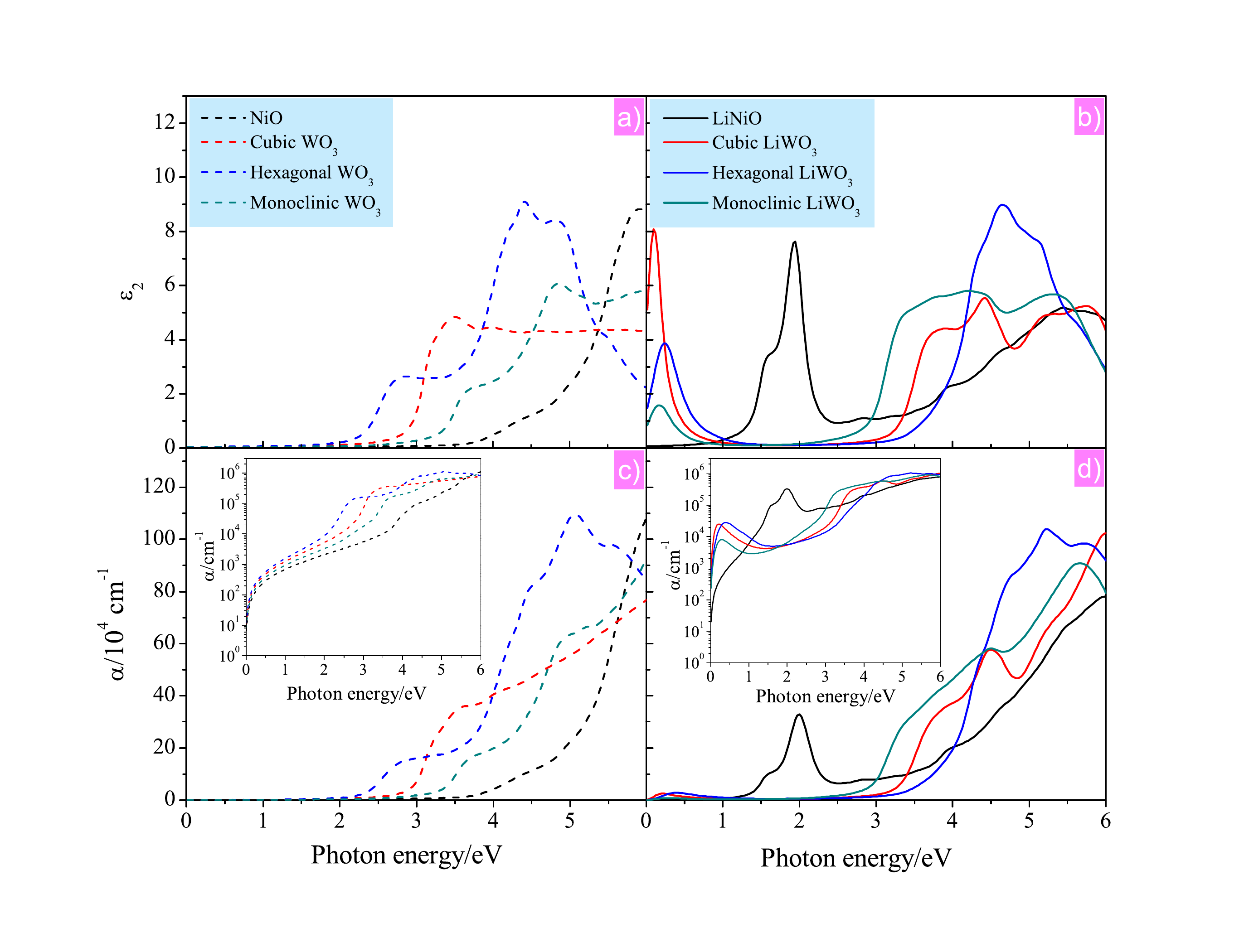}
\caption{Imaginary part of $\varepsilon_{yy}$ (a,b) as a function of photon energy for all systems. Absorption coefficient $\alpha$ (c,d). Insets show absorption coefficient on a logarithmic scale.}
\label{epsilon1}
\end{center}
\end{figure*} 

\subsubsection{Dielectric Response}
The optical response of materials to electromagnetic radiation is embedded in the dielectric tensor $\varepsilon_{ij}(\omega)$ which contains contributions from inter and intraband transitions. This tensor is symmetric with 6 independent components; the number of independent components depends on the symmetry under consideration. Cubic structures are highly symmetric and isotropic and only one component is independent. For the hexagonal symmetry only diagonal components appear; and for the monoclinic one the non-diagonal components can also be present \cite{cambrosch06a}. For our systems, the components $\varepsilon_{xx}$ and $\varepsilon_{yy}$ are comparable in magnitude and are the most intense of all components. The component $\varepsilon_{zz}$ in hexagonal and monoclinic symmetries is roughly half the magnitude of $\varepsilon_{xx}$; and the non-diagonal components (e.g., $\varepsilon_{xy}$) in the monoclinic structure of WO$_3$ are the least intense, about one order of magnitude smaller than the most intense ones. In addition, the imaginary part contains information related to inter and intraband transitions. Therefore, we shall focus only on the imaginary part of $\varepsilon_{yy}$ since this is the most intense in the near-infrared, visible, and ultraviolet regions for all systems. 

Figure \ref{epsilon1} (a,b) reports the imaginary part of $\varepsilon_{yy}$ (denoted simply as $\varepsilon_2$) as a function of photon energy for all systems. The results for the monoclinic phase of WO$_3$ are in good agreement with earlier theoretical and experimental reports \cite{johansson13a}. The onset of the dielectric constant for the pure symmetries follows, as expected, the same sequence as the magnitude of $E_g$, that is, in first place the hexagonal phase, then cubic, monoclinic and NiO. For energies below $E_g$ there are no features, indicating that the dielectric response for the pure systems is mainly dominated by interband transitions with photon energies exceeding $E_g$. These transitions are from O $2p$ states in the VB to W $5d$ and $6s$ states and Ni $3d$ and $4s$ states in the CB for WO$_3$ and NiO, respectively. Since electrons in the undoped systems are largely localized few intraband transitions are stimulated. For the doped structures of WO$_3$, free electrons are present. In these cases we observe a prominent peak in the far- and mid-infrared regions that is attributed to intraband transitions among $d$-$d$ states, then the curves fall down close to zero in the near-infrared and visible regions. The curves rise again first for monoclinic around 2 eV followed by the cubic and hexagonal, heading all to the ultraviolet region; all of these features are attributed to interband transitions. LiNiO exhibits no features in the infrared region, an indicative that no intraband transitions occur, however, a strong peak appears at 2.1 eV along with a small shoulder at 1.5 eV that can be assigned to intraband transitions (see below). Then the intensity falls down at the end of the visible region and goes up in the ultraviolet one, suggesting that in this region interband transitions take place. 

\subsubsection{Optical Absoprtion}
Having analyzed the dielectric function we now discuss its relationship with the absorption coefficient. According to section \ref{optteo} the dielectric function is related to the absorption coefficient through Equation \eqref{kii} and \eqref{alphaii}. The absorption in the visible region of the pure structures is largely dominated by the size of the energy gap. For photon energies lower than $E_g$, materials transmit most of the visible light since electrons in the VB cannot be promoted to the CB. Hence thin layers of NiO or WO$_3$ look transparent. Figure \ref{epsilon1} (c) shows that in the infrared region all systems exhibit the lowest absorption (inset shows the information in logarithmic scale for better appreciation). However, as the photon energy is increased beyond the absorption threshold, interband transitions take place. As it is pointed out above, the O 2$p$ states have a major weight in the VB close to the Fermi energy while the Ni $3d$ and W $5d$ states contribute largely to the CB and so interband transitions are allowed ($\Delta \ell =\pm1$). Note that the hexagonal phase has the highest absorption in most of the spectrum except for the region between 3 eV and 3.6 eV where is lower than the absorption of the cubic one. The absorption coefficient and the band gap are related by the well known Tauc's relation \cite{jtauc66a}: 
\begin{equation}
\label{tauceq}
\alpha E=A(E-E_g)^{\eta},
\end{equation}
where $A$ is an energy independent constant, $E$ the photon energy and $\eta$ is a power factor of the transition mode whose possible values are $\frac{1}{2}$, 2, $\frac{3}{2}$, and 3 for direct allowed, indirect allowed, direct forbidden, and indirect forbidden transition, respectively. According to the band structure calculations, the value of $\eta$ is 1/2 for monoclinic WO$_3$ and 2 for the other systems. Indirect transitions require the participation of phonons, so, it is expected that direct transitions dominate the absorption process. 

For the doped systems the situation is quite different. According to the DOS of LiWO$_3$, interband transitions are likely to occur from the B$_{-1}$ to the PFCB and from the PFCB to the B$_1$. In all cases, due to the Burstein-Moss shift, these transitions can only occur for photon energies larger than 2 eV. Figure \ref{epsilon1} (d) along with the inset shows that, indeed, the main absorption edge has slightly shifted to higher energies; about 0.3 eV for cubic and monoclinic structures and 0.6 eV for hexagonal symmetry, demonstrating that electronic transitions require larger excitation energies than in the pure structures. The monoclinic structure gives the largest absorption in the visible region, reaching its highest point in the blue-purple region with 2.7$\times10^5$ cm$^{-1}$. Since these systems are metallic it is expected that intraband transitions also partake in the absorption process \cite{cambrosch06a,mcazzaniga10}, particularly in the low energy region where an absorption peak appears for each structure. This feature is related to intraband transitions and agrees with previously published work within the context of polaron theory \cite{cyang14,cyang16,larsson03}.

For LiNiO the intricate nature of the band structure makes the absorption analysis more complicated due to the asymmetry in the band structure for both spin channels. Each spin channel makes asymmetric contributions to the absorption from both intra and interband transitions that makes difficult an accurate quantification. However, general trends can be drawn. In Figure \ref{epsilon1} (d) we observe that the onset of absorption has moved to lower energies and the magnitude of absorption has increased compared to that in NiO. Moreover, a prominent peak shows up at 2.1 eV reaching a maximum of 3.3$\times10^5$ cm$^{-1}$ and small shoulder at 1.5 eV. The main peak corresponds to absorption in the orangish region of the visible spectrum those transmitting a brownish color which is characteristic of the colored state of NiO. These features have been experimentally reported in Li$_x$NiO at 2.2 eV and 1.1 eV, respectively \cite{yzhang18}. Due to Li-insertion, free electrons are added to the system turning the oxide a half-metallic semimetal, so these features are associated with intraband transitions. 

\begin{figure}[t!]
\begin{center}
\includegraphics[width=9cm]{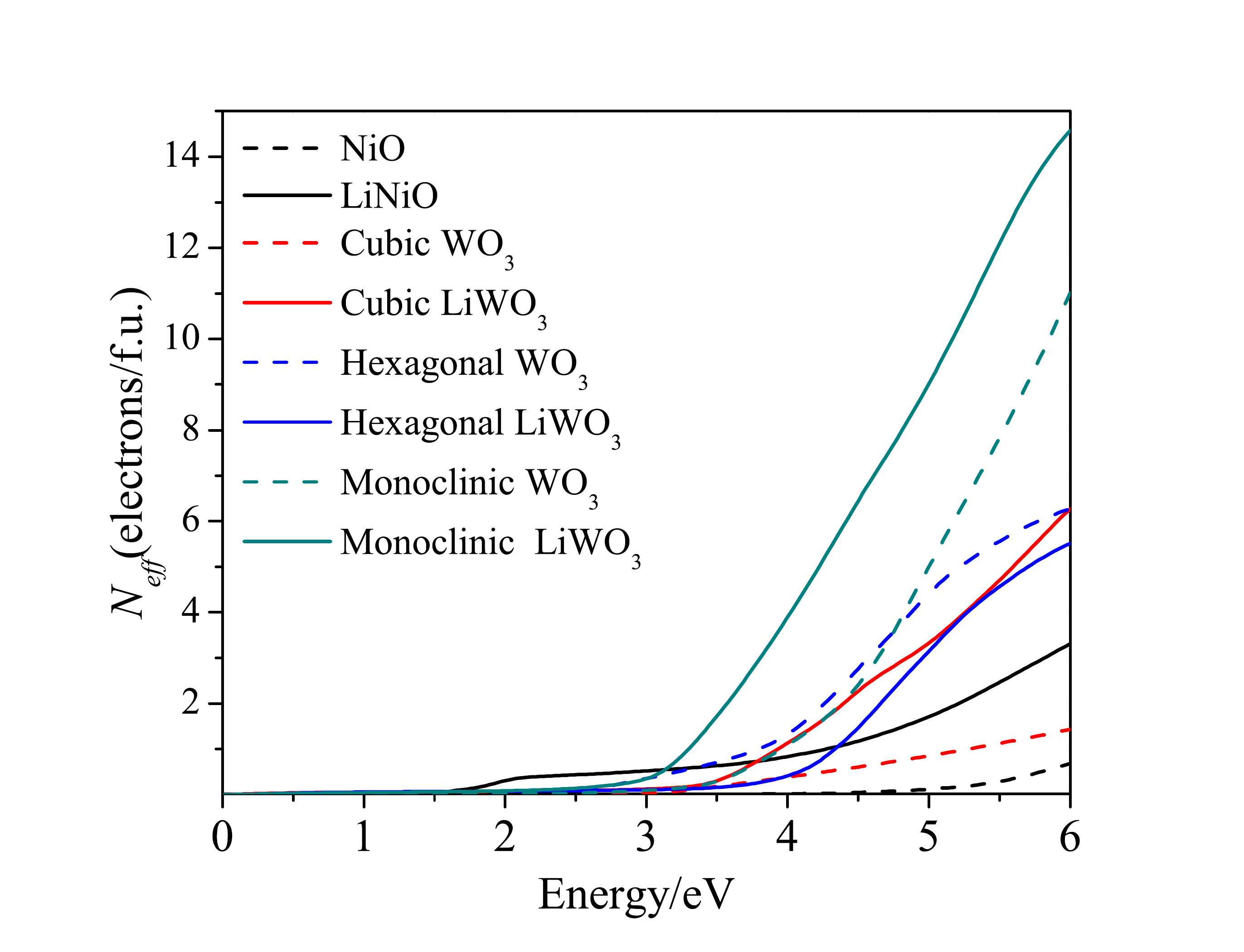}
\caption{Number of effective electrons for absorption}
\label{NefecNiOWO3}
\end{center}
\end{figure}

Lastly, Figure \ref{NefecNiOWO3} displays the number of effective electrons per formula unit during the absorption process as determined from the sum rule \eqref{sumr}. Overall, the extra electron added by the dopant fills the bottom of the conduction band and increases the Fermi level, increasing the number of electrons available for absorption. 
Note that NiO provides less than 1 electron at 6 eV and LiNiO about 3.3 at this same energy. For WO$_3$, the monoclinic undoped and doped phases give a maximum at 6 eV of 10.7 and 14.4 electrons, respectively. This is followed by the cubic undoped and doped phases with 1.5 and 6 electrons. The only exception is the hexagonal structure in which there is a slight reduction of electrons with respect to the pure phase. This may be attributed to the position of the Li atom since dopants in the hexagonal lattice can be accommodated in trigonal and/or hexagonal cavities. The latter is much larger and leaves more room for electron mobility \cite{pyang16a}. 

\section{Summary and conclusions}
\label{sec4}
We have explored from first principles calculations using the DFT+$U$ method the electronic and optical properties of Li-doped NiO and Li-doped WO$_3$; the latter for three symmetries: cubic, hexagonal and monoclinic. The results showed that the CB for the corresponding pure systems is mainly dominated by Ni $3d$ and W 5$d$ states whereas the VB has O $2p$ character. When Li is incorporated the Fermi level shifts to higher energies turning NiO a half-metallic semimetal and WO$_3$ a metal for the three symmetries. It is concluded that Li interstitial doping injects electrons to the system, driving the material to a metallic character, even for low concentrations of less than 8.3\%. It is found that monoclinic WO$_3$ exhibits a direct band gap, while the rest of the systems an indirect one. The nature of this gap in monoclinic WO$_3$ is still controversial and more detail experimental data are required keeping in mind not only its polymorphism at room temperature but also its anisotropic nature. Regarding the dielectric response, the onset of the imaginary part of the $\varepsilon_{yy}$ component for the pristine structures is first present for the hexagonal, then cubic, monoclinic and lastly for NiO in association with the magnitude of their band gap, an indication that interband transitions have the major contribution in the visible and ultraviolet regions. The hexagonal phase of pure WO$_3$ gives the maximum absorption in most of the energy interval, followed by cubic, monoclinic and NiO. It is demonstrated that, compared to the undoped systems, doping WO$_3$ with Li has the effect of increasing the absorption in the far- and mid-infrared regions while for LiNiO the effect is in the near-infrared and visible regions. LiNiO presents low absorption in the infrared region and a broad feature in the visible region attributed to intraband transitions. These findings shed light on the changes of the optical properties of both WO$_3$ and NiO with Li intercalation and can be of great interest for the designed of optimized ECD.

\section*{CRediT Author Statement}
Israel Perez: Conceptualization, Data curation, Formal analysis,
Investigation, Validation, Visualization, Writing - original draft, Project Administration, Supervision, Funding Acquisition. Juan Carlos Mart\'inez Faudoa: Conceptualization, Formal analysis, Investigation, Methodology, Software. Juan Ruben Abenuz Acu\~na: Conceptualization, Formal analysis, Investigation, Methodology, Software. Jose Trinidad Elizalde Galindo: Supervision, Project Administration, Funding Acquisition.

\section*{Declaration of competing interests} 
The authors declare that they have no known competing financial interests or personal relationships that could have appeared to influence the work reported in this paper.

\section*{Acknowledgements}
The authors thank the anonymous reviewers for critics that greatly improve the quality of this work. Funding: This work was supported by National Council of Science and Technology (CONACYT) Mexico under project 3035. We thankfully acknowledge the computer resources, technical advise, and support provided by Laboratorio Nacional de Superc\'omputo del Sureste de M\'exico (LNS), a member of the CONACYT network of national laboratories, Centro Nacional de Superc\'omputo (CNS), and Laboratorio Nacional de Inform\'atica (LANTI-UACJ). The authors are grateful to Oscar Ruiz for technical support.

\section*{Data Availability}
The raw/processed data required to reproduce these findings cannot be shared at this time as the data also forms part of an ongoing study.

\end{document}